\documentclass[]{aipproc}
\def\be{\begin{eqnarray}}
\def\ee{\end{eqnarray}}
\def\rme#1{\langle\mbox{#1}\rangle}
\def\bbox{\mathbf}
\def\vA{{\bbox  A}}
\def\vr{{\bbox  r}}
\def\vq{{\bbox  q}}
\def\vp{{\bbox  p}}
\def\vs{{\bbox \sigma}}
\def\vbp{{\overline \vp}}

\def\He#1{{}^{#1}\mbox{He}}

\def\nlo#1{\mbox{N$^{#1}$LO}}
\def\Sunit{\mbox{$10^{-20}$ keV-b}}
\def\dR{{\hat d^R}}

\layoutstyle{8x11single}
\begin{document}

%
%

\author{
{Tae-Sun Park}$^\ast$, {Kuniharu Kubodera}$^\ast$, {Dong-Pil Min}$^\dag$
and {Mannque Rho}$^{\dag\P\ddag}$}
{address={
 ${}^\ast$ Department of Physics and Astronomy,
University of South Carolina, Columbia, SC 29208, USA\\
 ${}^\dag$ School of Physics and Center for Theoretical Physics,
Seoul National University, Seoul 151-742, Korea\\
${}^{\P}$ Service de Physique Th\'{e}orique, CEA  Saclay,
\it
91191 Gif-sur-Yvette Cedex, France\\
${}^\ddag$ Institute of Physics and Applied Physics, Yonsei
University, Seoul 120-749, Korea}, }

\title{The Solar $pp$ and $hep$ Processes in Effective Field Theory}

\begin{abstract}
The strategy of modern effective field theory is exploited to pin
down accurately the flux $S$ factors for the $pp$ and $hep$
processes in the Sun. The technique used is to combine the high
accuracy established in few-nucleon systems of the ``standard
nuclear physics approach" (SNPA) and the systematic power counting
of chiral perturbation theory (ChPT) into a consistent effective
field theory framework. Using highly accurate wave functions
obtained in the SNPA and working to \nlo3 in the chiral counting
for the current, we make totally parameter-free and
error-controlled predictions for the $pp$ and $hep$ processes in
the Sun.
\end{abstract}

\maketitle \noindent In this talk, we report on the result of the
program sketched in \cite{strategy} which was made possible by the
collaboration with L.E. Marcucci, R. Schiavilla, M. Viviani, A.
Kievsky and S. Rosati, recently summarized in \cite{coll}.

\section{Introduction}

One of the ultimate goals of quantum chromodynamics (QCD) in
nuclear physics is to make precise model-independent
predictions for certain processes that figure importantly in
astrophysics and cosmology. The only technique presently available
to achieve such goals in the low-energy domain is effective field
theory (EFT) ~\cite{weinbergII}. In this talk, we apply a variant
of nuclear EFT developed by the authors to the solar $pp$ and
$hep$ processes within the framework sketched in \cite{strategy}.
In doing so, we rely on the accurate wave functions constructed by
Marcucci et al~\cite{MSVKRB} as the essential ingredient of EFT
coming from the standard nuclear physics approach (SNPA). The
power of the proposed scheme is the ability to correlate the beta
decay processes of $A=2, 3, 4$ nuclei that allows us to fix one
unknown constant in the theory, rendering possible a totally
parameter-free prediction of the two-nucleon and four-nucleon
processes.

The most abundant source of solar neutrinos (carrying  91 \% of
the total flux) is the $pp$ process \be p+p\rightarrow d + e^+ +
\nu\,. \ee This process has been carefully studied \cite{KB94, pp,
schiavilla, bc01}, and the calculated transition strength,
governed by the leading-order Gamow-Teller (GT) operator, is
believed to be reasonably reliable. However, given its extremely
important role in the solar burning process, the $pp$ process
invites further elaborate studies.
Meanwhile, the $hep$ process
\be
\He3+p\rightarrow \He4 + e^+ + \nu\,.
\ee
produces the highest energy
solar neutrinos, $E_\nu^{\rm max}(hep) = 20$~MeV.
While the $hep$ neutrino flux is estimated to be
much smaller than the
${}^8\mbox{B}$ neutrino flux,
there can be significant distortion of
the ${}^8\mbox{B}$ neutrino spectrum
at its higher end
if the $hep$ $S$-factor
is much larger than the existing estimates.
This change can influence the interpretation
of the results of a recent Super-Kamiokande experiment
that have raised many important issues
concerning the solar neutrino problem and neutrino
oscillations~\cite{controversy,monderen}.
To address these issues, a reliable estimate of the $hep$ cross
section is indispensable. 
Its accurate evaluation, however, has been
a long-standing challenge for nuclear and hadron
physics~\cite{challenge}. The difficulty involved is reflected in
the pronounced variance in the documented estimates of the $hep$
$S$-factor. For example, the first estimate given by
Salpeter~\cite{salpeter} was $S(hep)= 630\times \Sunit$, which was
eventually replaced by the much smaller (so-called ``standard'')
value, $\simeq 2\times\Sunit$~\cite{CRSW91,SWPC92}; the latest,
most elaborate estimation gives $9.64\times \Sunit$~\cite{MSVKRB}.
The reason for the difficulty in making a precise estimation of
the $hep$ $S$-factor is multifold. First, the one-body (1B) GT
matrix element for the $hep$ process is strongly suppressed due to
the symmetry properties of the orbital wave functions of the
initial and final states. The main orbital wave function for
$\He4$ has [4] symmetry (totally symmetric) under the particle
exchanges, whereas the dominant $\He3+p$ orbital wavefunction has
[31] symmetry; the [4] component is forbidden by the Pauli
principle when there are three protons. Then, the main components
of the initial and final states -- with different symmetry
properties in orbital space -- cannot be connected by the
leading-order (LO) GT operator, which does not contain orbital
variables. This means that the non-vanishing 1B GT matrix element
(for the $hep$ process) is due to either minor components of the
wavefunctions or higher order corrections of the GT operators.
Since these are all quite small, the 1B matrix element becomes
comparable to multi-body corrections, e.g., meson-exchange-current
(MEC) contributions. A further complicating feature is that there
is a substantial cancellation between the 1B and two-body (MEC)
contributions, which can amplify the errors. Finally, with the
``chiral-filter mechanism" rendered ineffective, at non-vanishing
leading order of effective field theory (see below), the many-body
corrections contain short-ranged operators the strengths of which
are not known {\it a priori} and hence difficult to control.

The objective of our present work is to make accurate effective
field theory (EFT) predictions on the $pp$ and $hep$ processes
within a single framework. For this purpose, we adopt the strategy
that exploits the power of {\it both} SNPA and heavy-baryon chiral
perturbation theory (HBChPT), which is a well-studied low-energy
EFT. In this approach, EFT enters in the calculation of relevant
transition operators. We will calculate them up to
next-to-next-to-next-to-leading order (\nlo3); all the operators
that appear up to \nlo3 will be considered. To obtain the
corresponding nuclear matrix elements, we need highly accurate
nuclear wave functions. Although it is- at least in principle --
possible to also derive nuclear wave functions to the appropriate
order from HBChPT, we choose not to do so. Instead, we use
realistic wave functions obtained in the standard nuclear physics
approach (SNPA). The potentials that generate such wave functions
are supposed to contain high orders in the chiral counting,
presumably much higher than what can be accounted for in the
irreducible vertex for the current operators. For a review of
SNPA, see Ref.~\cite{snpa}.  Such an approach-- which is close in
spirit to Weinberg's original scheme~\cite{weinberg} based on the
chiral expansion of ``irreducible terms"--has been found to have
an amazing predictive power for the $n + p\rightarrow d+\gamma$
process~\cite{np,npp,PNC}.

The basic advantage of EFT is that the SNPA and HBChPT can be
combined into a model-independent framework based on the first
principle. A systematic expansion scheme of EFT reveals that, for
the GT transition for which the ``chiral-filter
mechanism"~\cite{strategy} is rendered inoperative, the
one-pion-exchange (OPE) operators and the leading short-ranged
operators have the same chiral order, and hence their
contributions should be comparable. This nullifies the intuitive
argument that the long-ranged OPE contribution should dominate the
MEC corrections. The existing SNPA calculation, however, lacks
this leading short-ranged operator while containing instead some
higher order short-ranged contributions. What our EFT manages to
do is to account for this short-ranged contribution in a way
consistent with renormalization group invariance. This is a novel
way of understanding the so-called ``short-range correlation" in
SNPA. We will see that this aspect indeed plays an essential role
both for $pp$ and $hep$ processes.

Briefly, our approach to HBChPT is as follows. We take only pions
and nucleons as pertinent degrees of freedom. All others have been
integrated out, and their dynamical roles are embedded in the
higher-order operators.
In the scheme relevant to us, it suffices to focus on
``irreducible graphs" according to Weinberg's
classification~\cite{weinberg}. Graphs are classified by the
chiral power index $\nu$ given by $\nu = 2 (A-C) + 2 L +\sum_i
\nu_i \label{nu}$, where $A$ is the number of nucleons involved in
the process, $C$ the number of disconnected parts, and $L$ the
number of loops. The chiral index, $\nu_i$, of the $i$-th vertex
is given by $\nu_i= d_i + e_i + n_i/2-2 \label{nui}$, where $d_i$,
$e_i$ and $n_i$ are respectively the numbers of derivatives,
external fields and nucleon lines belonging to the vertex. The
Feynman diagrams with a chiral index $\nu$ are suppressed by
$(Q/\Lambda_\chi)^\nu$ compared with the leading-order one-body GT
operator, with $Q$ standing for the typical three-momentum scale
and/or the pion mass, and $\Lambda_\chi \sim m_N \sim 4\pi f_\pi$
is the chiral scale. The physical amplitude is then expanded with
respect to $\nu$.

\section{Theory: GT operators up to \nlo3}
The LO and \nlo2 contributions come from the well-known
one-body currents;
the GT operator for the $a$-th isospin component reads
\be
\vA^a_{\rm 1B} = g_A \sum_{l=1,2}
\frac{\tau^a_l}{2} \left[
 \vs_l + \frac{\vbp_l\, \vs_l
 \cdot \vbp_l - \vs_l \, {\bar p}_l^2}{2 m_N^2}
 \right],
 \label{1-body}
\ee
with $\vbp_l\equiv (\vp_l+ \vp_l^{\,\prime})/2$
and $g_A\simeq 1.267$.
Corrections to the above 1B operators are due to MECs,
which start at \nlo3.
In our work we include {\it all}
the contributions up to \nlo3.
We emphasize in particular
that, up to \nlo3, only two-body (2B) currents enter,
three-body currents appearing only from \nlo4;
thus there is no arbitrary truncation involved here.
The \nlo3 2B currents consist of
the OPE and contact-term (CT) parts,
\be
\vA_{{\rm 2B}}^a =
\vA_{{\rm 2B}}^a(\mbox{OPE}) +
\vA_{{\rm 2B}}^a(\mbox{CT}).
\ee
The OPE part is given as
\be
\vA_{{\rm 2B}}^a(\mbox{OPE}) &=&
- \frac{g_A}{2 m_N f_\pi^2}\, \frac{1}{m_\pi^2 + q^2}
 \Bigg[ - \frac{i}{2} \tau^a_\times\,\vp\,\,
     \vs_-\cdot\vq
 + 2 \,\hat c_3\, \vq \,\, \vq\cdot
  (\tau_1^a \vs_1 +\tau_2^a \vs_2)
 + \left(\hat c_4
  + \frac14\right) \tau^a_\times\,
 \vq \times \left[ \vs_\times \times \vq\, \right]
 \frac{}{}\Bigg]\,,
\ee
where $\tau^a_{\odot}\equiv (\tau_1 \odot \tau_2)^a$,
with $\odot=\times,-$,
and similarly for $\vs_\odot$.
Since the couplings ${\hat c}_{3,4}$ are determined from the
$\pi N$ data~\cite{csTREE},
$\hat c_3 = -3.66 \pm 0.08$
and $\hat c_4 = 2.11 \pm 0.08$,
there is no unknown parameter here.
Now the CT part is given as
\be
\vA_{{\rm 2B}}^a(\mbox{CT}) =
 - \frac{g_A}{m_N f_\pi^2} \left[
  \hat d_1 (\tau_1^a \vs_1 + \tau_2^a \vs_2)
 + \hat d_2 \tau^a_\times \vs_\times
 \right] ,\label{2-body}
\ee
which contains two low-energy constants.
However, the Pauli principle
(or the ``$L+S+T=\mbox{odd}$'' rule)
combined with the fact
that the CT term is effective only for $s$-wave ($L=0$)
implies that
we need only work with one unknown constant,
$\dR$, defined by
\be
 \dR\equiv \hat d_1 +2 \hat d_2 +
\frac13 \hat c_3
 + \frac23 \hat c_4 + \frac16 \>\>.
\label{dr} \ee Furthermore, the same combination also enters into
tritium $\beta$-decay, $\mu$-capture on deuteron, and $\nu$--$d$
scattering. Although $\dR$ is in principle calculable from QCD for
a given scale $\Lambda$ , this calculation is not available at
present; we therefore need to fix $\dR$ empirically. Here we
choose to determine $\dR$ by fitting the tritium $\beta$-decay
rate, $\Gamma_\beta$, which is accurately known
experimentally~\cite{schiavilla}. Once $\dR$ is fixed, our
calculation involves no unknown parameters.

We calculate the matrix elements of the transition operators
with state-of-the-art
realistic nuclear wave functions for $A=2,\,3,\,4$.
We employ the correlated-hyperspherical-harmonics (CHH)
wave functions, obtained with the
Argonne $v_{18}$ (Av18) potential
(supplemented with the Urbana-IX three-nucleon
potential for the $A\ge 3$ nuclei).
To control short-range physics in a consistent manner,
we apply the same regularization method
to all the nuclear systems in question.
Specifically, in performing Fourier transformation
to derive the $r$-space representation of
transition operators,
we use the Gaussian regularization.
This is equivalent
to replacing the delta and Yukawa functions
with the regularized ones,
$$
\left(\delta_\Lambda^{(3)}(r),\, y_{0\Lambda}^\pi(r) \right)
 \equiv
 \int \!\!\frac{d^3 q}{(2\pi)^3}\,
  S_\Lambda^2(q^2)\, e^{ i \vq\cdot \vr}
  \left(1,\,\frac{1}{q^2 + m_\pi^2}\right) ,
$$
where the cut-off function
$S_\Lambda(q^2)$ is defined as
\be S_\Lambda(q^2) =
\exp\left(-\frac{q^2}{2\Lambda^2}\right). \label{regulator} \ee
The cutoff parameter $\Lambda$ characterizes the energy-momentum
scale of our EFT.

\section{Results}
The value of $\dR$ determined
from the experimental value of $\Gamma_\beta$ is
\be
\dR=(1.00 \pm 0.07,\ 1.78 \pm 0.08,\ 3.90 \pm 0.10)
\ee
for the choice of $\Lambda=(500,\ 600,\ 800)$ MeV,
respectively.
We list in Table~\ref{table:pp}
the GT matrix elements
for the $pp$ and $hep$ processes (in arbitrary units)
as a function of $\Lambda$.
\begin{table}[ht]
\begin{tabular}{c|cc|cc}\hline
$\Lambda$ (MeV) & $\rme{1B}_{pp}$ & $\rme{2B}_{pp}$  
                & $\rme{1B}_{hep}$ & $\rme{2B}_{hep}$\\
\hline
500 & 4.82 &
$0.076 - 0.035\ \dR \simeq 0.041$
& $-0.81$ &
$0.93 -0.44\ \dR \simeq 0.49$
\\
600 & 4.82 &
$0.097 - 0.031\ \dR \simeq 0.042$
 & $-0.81$ &
$1.22 -0.39\ \dR \simeq 0.52$ 
\\ 
800 & 4.82 &
$0.129 - 0.022\ \dR \simeq 0.042$
& $-0.81$ &
$1.66 - 0.27\ \dR \simeq 0.59$
\\ \hline
\end{tabular}
\caption{\label{table:pp}\protect
GT matrix element for the $pp$ and $hep$ processes,
calculated for representative values of $\Lambda$.
The 2B contribution is the sum of the
OPE part ($\dR$-independent)
and the CT part (linear in $\dR$), for each case.}
\end{table}
\subsection{The $pp$ results}
We observe that, while the OPE part by itself has
a sizable $\Lambda$-dependence,
the net amplitude is completely $\Lambda$-independent.
In other words, the $\Lambda$-dependence of the OPE part
has been perfectly removed by that of CT part.
The
$\Lambda$-independence of the physical quantity,
$\rme{1B}+\rme{2B}$,
which is
in conformity with the general {\it tenet} of EFT,
is a crucial feature
of the result in our present study.

The relative strength of the
two-body contribution as compared
with the one-body contribution is
\be
\delta_{\rm 2B}^{pp} \equiv \rme{2B}_{pp} /\rme{1B}_{pp}
= (0.86 \pm 0.05)\ \%.
\ee
This ratio is consistent with the latest
 SNPA calculation~\cite{schiavilla},
$\delta_{\rm 2B}^{pp} = (0.5 \sim 0.8)\ \%$.
The resulting $pp$ $S$-factor is
\be S(pp)
= 3.94\ (1 \pm 0.15\ \% \pm 0.1\ \%) \times 10^{-25}\ \mbox{MeV-barn}\,,
\label{S-factor}
\ee
where
the first and the second uncertainties
come from one- and two-body contributions,
respectively.

\subsection{The $hep$ results}
The general tendency here
is quite similar to the $pp$ case;
the variation of the two-body GT amplitude
is only $\sim$10 \% for the entire range
of $\Lambda$ under study.
The $\Lambda$-dependence in the total GT amplitude becomes
more pronounced by a strong cancellation between the 1B
and 2B terms, but this amplified $\Lambda$-dependence still
lies within acceptable levels.

\begin{table}[hbt]
\centering
\begin{tabular}{c|ccc|c}\hline
$\Lambda$ (MeV) & 500 & 600 & 800 & MSVKRB
\\ \hline
${}^1S_0$ & 0.02  &  0.02  &  0.02 & 0.02 \\
${}^3S_1$ & 7.00  &  6.37  &  4.30 & 6.38 \\
${}^3P_0$ & 0.67  &  0.66  &  0.66 & 0.82 \\
${}^1P_1$ & 0.85  &  0.88  &  0.91 & 1.00 \\
${}^3P_1$ & 0.34  &  0.34  &  0.34 & 0.30 \\
${}^3P_2$ & 1.06  &  1.06  &  1.06 & 0.97 \\ \hline
Total     & 9.95  &  9.37  &  7.32 & 9.64 \\ \hline
\end{tabular}
\caption{\label{TabS}\protect Contributions
to the $S$-factor (in \Sunit)
from individual initial channels calculated as functions of $\Lambda$.
The last column gives the results obtained in MSVKRB.}
\end{table}

Table~\ref{TabS} shows the contribution to the $S$-factor
from each initial channel, at zero c.m. energy.
For comparison we have also listed the
latest results based on SNPA
\cite{MSVKRB} (which we refer to as MSVKRB).
It is noteworthy that for all the channels
other than ${}^3S_1$,
the $\Lambda$-dependence is very small ($<2$ \%).
While the GT terms are dominant,
the contribution of the axial-charge term
in the ${}^3S_1$ channel is sizable
even though it is kinematically suppressed
by the factor $q$.

The results given in Table~\ref{TabS}
lead to a much improved estimate
of the $hep$ $S$-factor:
\be
S(hep)=(8.6 \pm 1.3 )\times \Sunit\,,\label{prediction}
\ee
where the ``error"
spans the range of the $\Lambda$-dependence
for $\Lambda$ = 500--800 MeV.
This result is to be compared to that obtained by
MSVKRB~\cite{MSVKRB}, $S=9.64 \times \Sunit$.
To decrease the
uncertainty in Eq.(\ref{prediction}), we need to reduce the
$\Lambda$-dependence in the two-body GT term. According to a
general {\it tenet} of EFT, the $\Lambda$-dependence should
diminish when higher order terms are included.
A preliminary
study indicates that it is indeed possible to
reduce the $\Lambda$-dependence significantly by including \nlo4
corrections.

\section{Discussion}
By determining the only parameter of the theory $\dR$ from the
experimental data on triton beta decay, we have succeeded in
making rather accurate EFT predictions (up to \nlo3) in a
parameter-free manner for both $pp$ and $hep$ processes. These
predictions turn out to give support to the latest SNPA results.

The prediction for the $pp$ prediction comes out to be independent
of the cutoff scale $\Lambda$, which means that it is fully
consistent with the tenet of EFT. On the other hand, there remains
some $\Lambda$-dependence for the $hep$ process, which could be due
to many-body nature absent in the $pp$ case. Even so, it is
remarkable that the theoretical uncertainty can be reduced 
from ``orders of magnitude'' to $\sim 20$ \%. 
This uncertainty can be
further reduced if \nlo4 terms -- which involve no additional
unknowns -- are taken into account.

We should note that by using the ``exact" wave functions, we are
sacrificing the strict adherence to chiral order counting in favor
of predictivity. The counting error committed therein comes at one
order higher than that accounted for in the irreducible vertex for
the current , i.e., at \nlo4 in the present calculation, and
should be small for the whole scheme to make sense. This can be
checked by an \nlo4 calculation for which the counting error would
come at \nlo5 or higher. Furthermore the notion implicit in our
approach that the 1B matrix element calculated with the ``exact"
wave functions should be taken as ``empirical" could be checked by
looking at the $hen$ process $$\He3+n\rightarrow \He4 + \gamma$$
in which the 1B matrix element is expected to suffer the same
suppression due to the symmetry properties of the initial and
final states. 

It seems likely that an EFT calculation of the $hep$ process that
adheres to the strict power counting -- such as the pionless
theory~\cite{beaneetal} -- would be obstructed by a plethora of
unknown parameters that are difficult to completely control. If
such a calculation is feasible, however, it would be interesting
to see if and how the symmetry suppression and the sensitive
cancellation encountered in our version of EFT -- where an
accurate 1B matrix element plays a key role -- could either be
circumvented or manifest themselves in the description.

\end{document}